\documentstyle[11pt,newpasp,twoside,epsf]{article}
\def\edcomment#1{\iffalse\marginpar{\raggedright\sl#1\/}\else\relax\fi}
\reversemarginpar

\begin{document}
\title{Protoplanetary disk in V645 Cyg as seen with H$_2$O and methanol masers.} 
\author{I.E. Val'tts}
\affil{Astro Space Center, Profsouznaya st. 84/32, 117997 Moscow, Russia}
\author{V.I. Slysh}
\affil{Astro Space Center, Profsouznaya st. 84/32, 117997 Moscow, Russia;
Arecibo Observatory, HC 3 Box 53995, Arecibo PR 00612, USA}
\author{M.A. Voronkov}
\affil{Astro Space Center, Profsouznaya st. 84/32, 117997 Moscow, Russia}
\author{V. Migenes}
\affil{University of Guanajuato, Department of Astronomy, Apdo
Postal 144, Guanajuato, CP36000, GTO, Mexico}

\begin{abstract}
Radio images of maser spots in the infrared source 
GL2789, connected with the young stellar object V645 Cyg, 
have been obtained as a result of the radio interferometric 
observations of H$_2$O maser at 22 GHz and  methanol maser 
at 6.7 GHz, with the VLBI arrays VLBA and EVN. It was 
shown that the position of the masers coincide with the 
optical object within 0\farcs2. The maser spots are located 
along the line North-South, and their position and radial 
velocity can be described by a model of the Keplerian disk 
with a maximum radius of 40 AU for H$_2$O maser and 800 AU 
for methanol maser. The H$_2$O and methanol maser spots 
have not been resolved, and lower limits of the brightness 
temperature is $2\times10^{13}$~K and $1.4\times10^9$~K,
respectively. A model of the maser was suggested in which
the maser emission is generated in extended water and methanol 
envelopes of icy planets orbiting the young star.
\end{abstract}

\section{Introduction}

GL2789 is an infrared source coinciding with the 
optical reflection nebula and star-like object V645 Cyg 
(Cohen 1977). On the optical image there is a star-like 
condensation N0 and several filamentary nebulae, the 
brightest of them named N1. Cohen (1977) proposed that 
this object is connected with a young  O7 star, at the 
distance 6 kpc (3.5 kpc - in Gudrich 1986) with a mass 
10M$_\odot$. V645Cyg is located at the center of molecular cloud 
emitting in CO and NH$_3$ molecular lines (bipolar outflow in 
CO: Schulz et al. 1989, Torrelles et al. 1987), and at the 
center of thermal continuum radio source of the size  7\arcsec
(Skinner et al. 1993). Lada et al. (1981) has found H$_2$O 
maser at two velocities -48.9 km~s$^{-1}$ and -44.5 km~s$^{-1}$, 
coinciding in position with the optical object V645Cyg 
within 0\farcs2. 13 years later Tofani et al. (1995) has observed 
this  maser  with  the  VLA at the velocities  $-$43.3 km~s$^{-1}$ and
$-$41.0 km~s$^{-1}$ at the same position. OH maser emission at the 
frequency 1665 MHz was found by Morris and Kazes (1982) 
in  the  interval of  the  radial velocities  from  -45 km~s$^{-1}$ to  
-41.6 km~s$^{-1}$. Slysh et al. (1999) found a methanol maser at  
6.7 GHz in the transition $5_1-6_0$~A$^+$ in the interval of  
velocities from $-$43.5 km~s$^{-1}$ to $-$40.5 km~s$^{-1}$. 
Since maser emission is associated with early stages of 
stellar evolution, V645 Cyg must be a protostar or very 
young star.  Study of the fine structure of maser sources, 
connected with V645 Cyg, could help to understand a 
mechanism of the interaction between young star or 
protostar and surrounding matter. 
In this presentation we give results of the VLBI study 
of H$_2$O and methanol masers with high (milliarcsec) 
angular resolution.

\section{Observations}
The H$_2$O observations were made on June 6, 1996 as 
part of a 24-hour VLBA survey of potential VSOP targets of 
continuum sources at 5 GHz and H$_2$O maser targets 
(Migenes et al. 1999).  The source GL2789 was observed at 
22.235 GHz for 5 minutes and the polarization was left-hand circular.
The data reduction was made using the AIPS 
software system and details are given by Migenes et al. 
(1999).  The synthesized beam was 1.0$\times$0.3 mas, with the 
position angle 7\arcsec.  The bandwidth of 8 MHz was divided 
into 512 spectral channels, providing spectral resolution 
15.6~kHz per channel or 0.21~km~s$^{-1}$.  The noise of the 
cleaned image corresponded to the weakest details of 
about 3 Jy~beam$^{-1}$ (3$\sigma$ level).

Methanol  observations  were made in the transition 
$5_1-6_0$~A$^+$ at the frequency 6.7~GHz with the EVN (European 
VLBI) in 1998 and 2000. 5 telescopes have taken part: 
Effelsberg 100m, Jodrell Bank 25m, Medicina 32m, 
Onsala 25m, Torun 32m. In 1998 GL2789 was observed 
during 4 10-minutes intervals, in 2000 during 3.  The 
synthesized beam  was  4.3$\times$8.2 mas.  The bandwidth  of  
2 MHz was devided into 1024 spectral channels, providing 
spectral resolution 1.95~kHz per channel or 0.088~km~s$^{-1}$. 
The noise of the cleaned image corresponded to the 
weakest details of about 3 Jy~beam$^{-1}$ (3$\sigma$). The
post-processing was done in standard VLBA manner (Diamond  
1995) in AIPS package: the amplitude calibration, fringe 
fitting, determination of absolute position and mapping of 
the maser spots.

Relative positions of H$_2$O and methanol maser spots
were given in Table~1. Fig.~1 shows positions of H$_2$O maser spots
relatively the reference feature at the velocity $-$51.8~km~s$^{-1}$ and
spectra of both methanol and water vapor masers. The maps of methanol
masers obtained for two epochs are shown in Fig.~2. The map of H$_2$O
maser as well as the relative positions of methanol and H$_2$O masers
are shown in Fig.3. Finally, Fig.~4. presents a velocity-position diagram
for masers in both molecules and the scatch of proposed disk model
explaining observed velocity and positional properties of these masers.

\begin{table}
\caption{Relative positions of H$_2$O and CH$_3$OH maser spots.}
\label{relpos}
\vskip 1mm
\begin{tabular}{ccrrrrrr}
\hline
\multicolumn{2}{c}{\Large\strut}&\multicolumn{3}{c}{CH$_3$OH (1998)}&\multicolumn{3}{c}{CH$_3$OH (2000)}\\
\hline
Feature&\multicolumn{1}{c}{V$_{\mathrm{LSR}}$}&\multicolumn{1}{c}{$\Delta\alpha cos\delta$}&\multicolumn{1}{c}{$\Delta\delta$}&\multicolumn{1}{c}{Flux}&\multicolumn{1}{c}{$\Delta\alpha cos\delta$}&\multicolumn{1}{c}{$\Delta\delta$}&\multicolumn{1}{c}{Flux}\\
&&&&\multicolumn{1}{c}{density}&&&\multicolumn{1}{c}{density}\\
&\multicolumn{1}{c}{km~s$^{-1}$}&\multicolumn{1}{c}{mas}&\multicolumn{1}{c}{mas}&\multicolumn{1}{c}{Jy}&\multicolumn{1}{c}{mas}&\multicolumn{1}{c}{mas}&\multicolumn{1}{c}{Jy}\\
\hline
A&$-$43.4 &  76.0 & $-$45.1 &  2.0 & 77.7 & $-$45.4 &  2.5 \\
B&$-$42.5 &  85.0 & $-$65.0 &  2.5 &  83.5 & $-$64.4 &  1.2 \\
B\makebox[0pt]{$_{~\;1}^{~~*}$}&$-$42.8&83.7&$-$63.2& 1.0& 82.8 & $-$64.7 &  0.9 \\
C&$-$40.8 &  82.9 & $-$134.2 &  2.0 & 82.8 & $-$134.5 &  1.7 \\
D&$-$40.4 &  94.2 & $-$143.2 &  2.2 & 94.1 & $-$143.3 &  3.0 \\
\hline
\multicolumn{2}{c}{\Large\strut}&\multicolumn{3}{c}{H$_2$O (1996)}&\multicolumn{3}{c}{}\\
\hline
&$-$54.3 &     0.40 &    0.22 &  4.7 &\multicolumn{3}{c}{}\\
&$-$51.8 &     0.03 &    0.00 & 36.0 &\multicolumn{3}{c}{}\\
&$-$50.5 &  $-$3.76 & $-$4.16 &  1.8 &\multicolumn{3}{c}{}\\
&$-$49.5 &  $-$0.27 & $-$0.04 & 10.0 &\multicolumn{3}{c}{}\\
&$-$48.2 &  $-$0.80 & $-$0.06 &  0.6 &\multicolumn{3}{c}{}\\
&$-$41.6 &     0.38 &    8.10 &  6.6 &\multicolumn{3}{c}{}\\
&$-$41.0 &     0.40 &    8.04 &  6.4 &\multicolumn{3}{c}{}\\
&$-$39.0 &     0.45 &    7.71 &  7.3 &\multicolumn{3}{c}{}\\
\hline
\end{tabular}
\vskip -3mm
\par\noindent
\parbox{\linewidth}{\flushleft
$^{*}$~-- the feature corresponds to wings of the line at
$-$42.5~km~s$^{-1}$.\hfill
}
\end{table} 

\section{Discussion and conclusion}
\begin{figure}
\plottwo{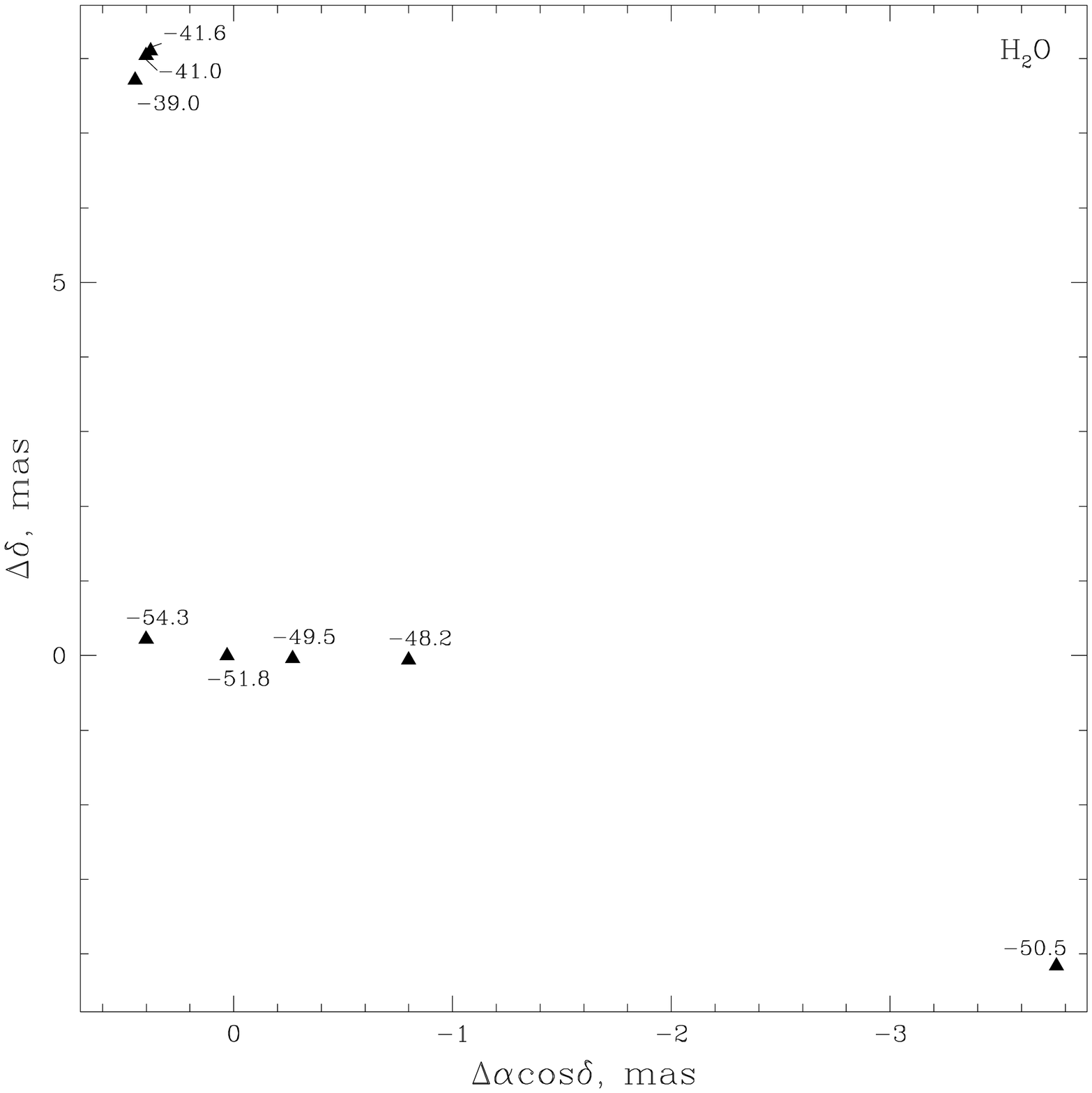}{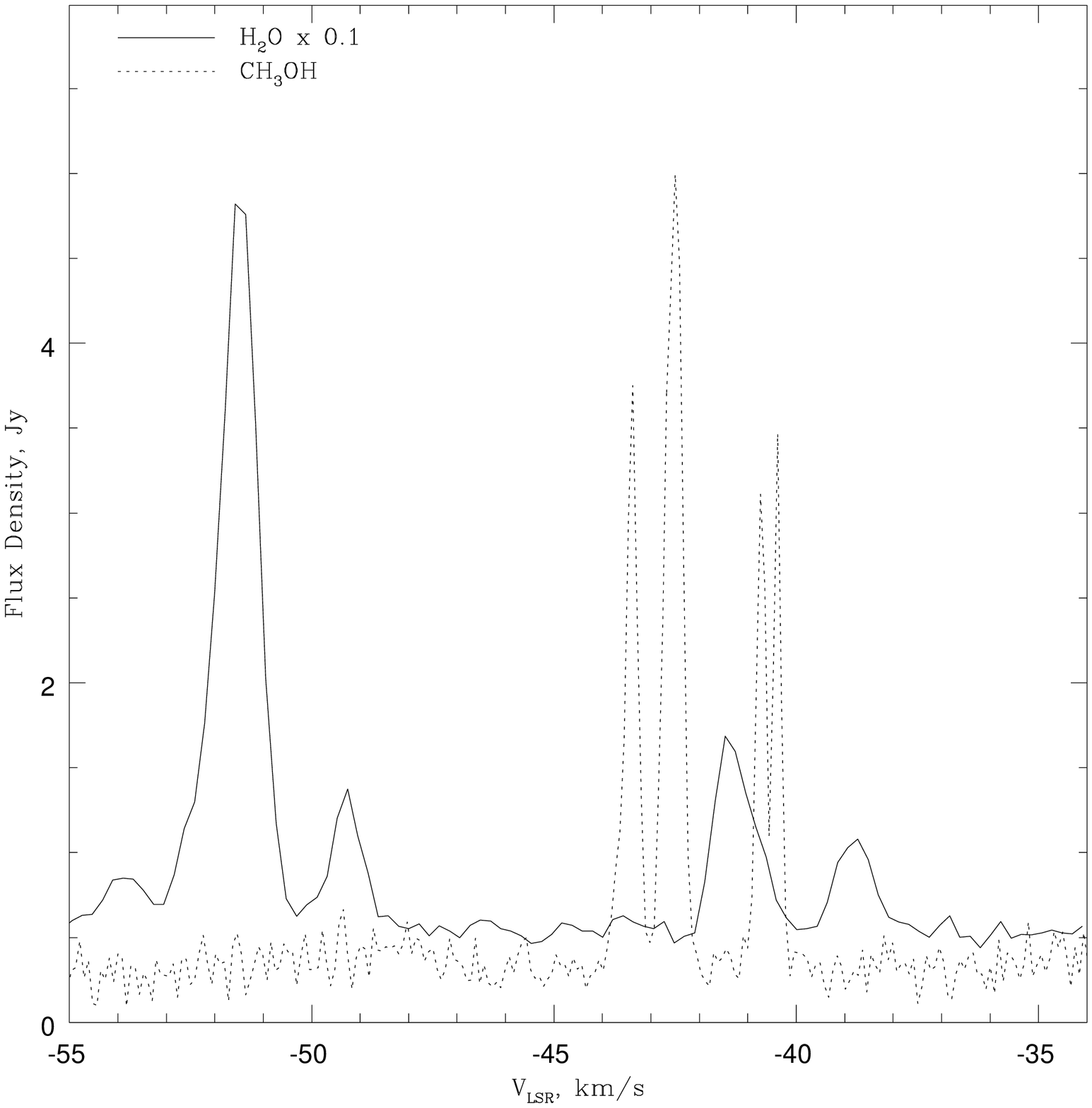}
\caption{Positions of the H$_2$O maser spots relatively the reference
feature at the velocity $-$51.8~km~s$^{-1}$ (Left). Spectra of H$_2$O
(solid line) and CH$_3$OH (dashed line) masers (Right).}
\end{figure}

\begin{figure}
\plottwo{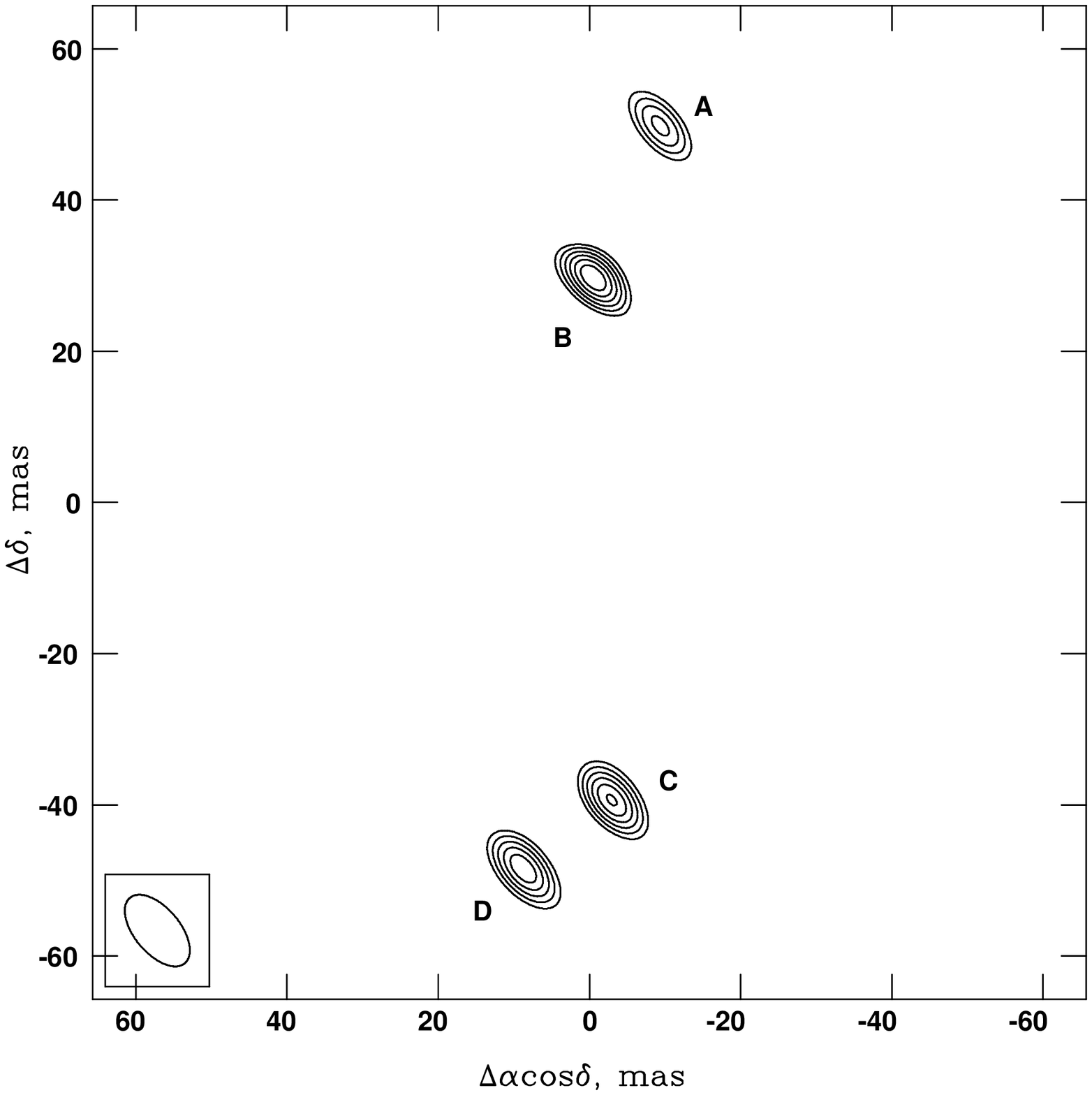}{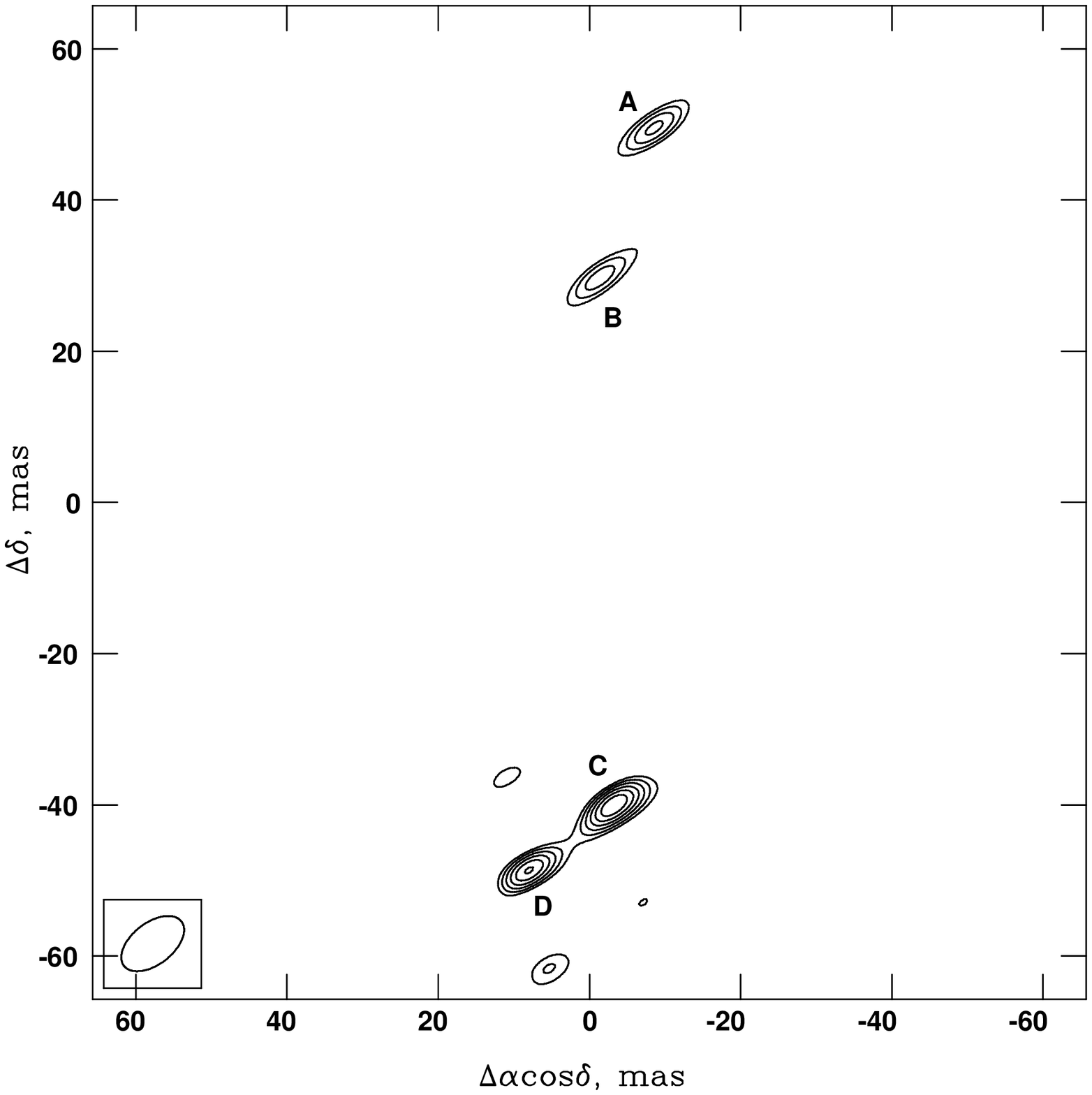}
\caption{Map of methanol maser. Left figure is observations of 1998,
contours are 0.22$\times$(4, 5, 6, 7, 8, 9)~Jy/beam. Right figure is observations
of 2000, contours are 0.35$\times$(4, 5, 6, 7, 8, 9)~Jy/beam.
A, B, C, D -- components at velocities $-$43.4, $-$42.5, $-$40.7 and
$-$40.4 km~s$^{-1}$, respectively.}
\end{figure}

\begin{figure}
\plottwo{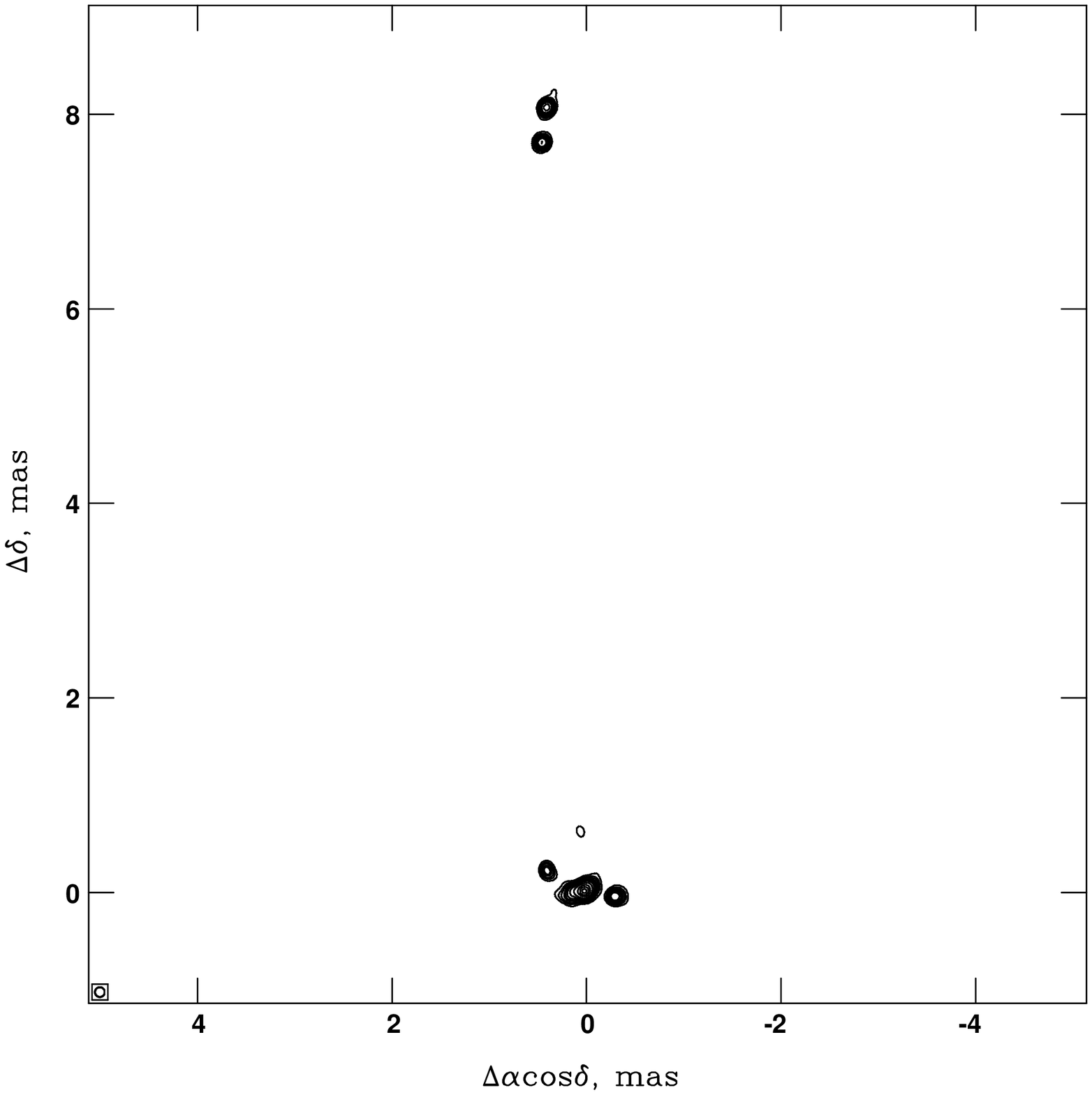}{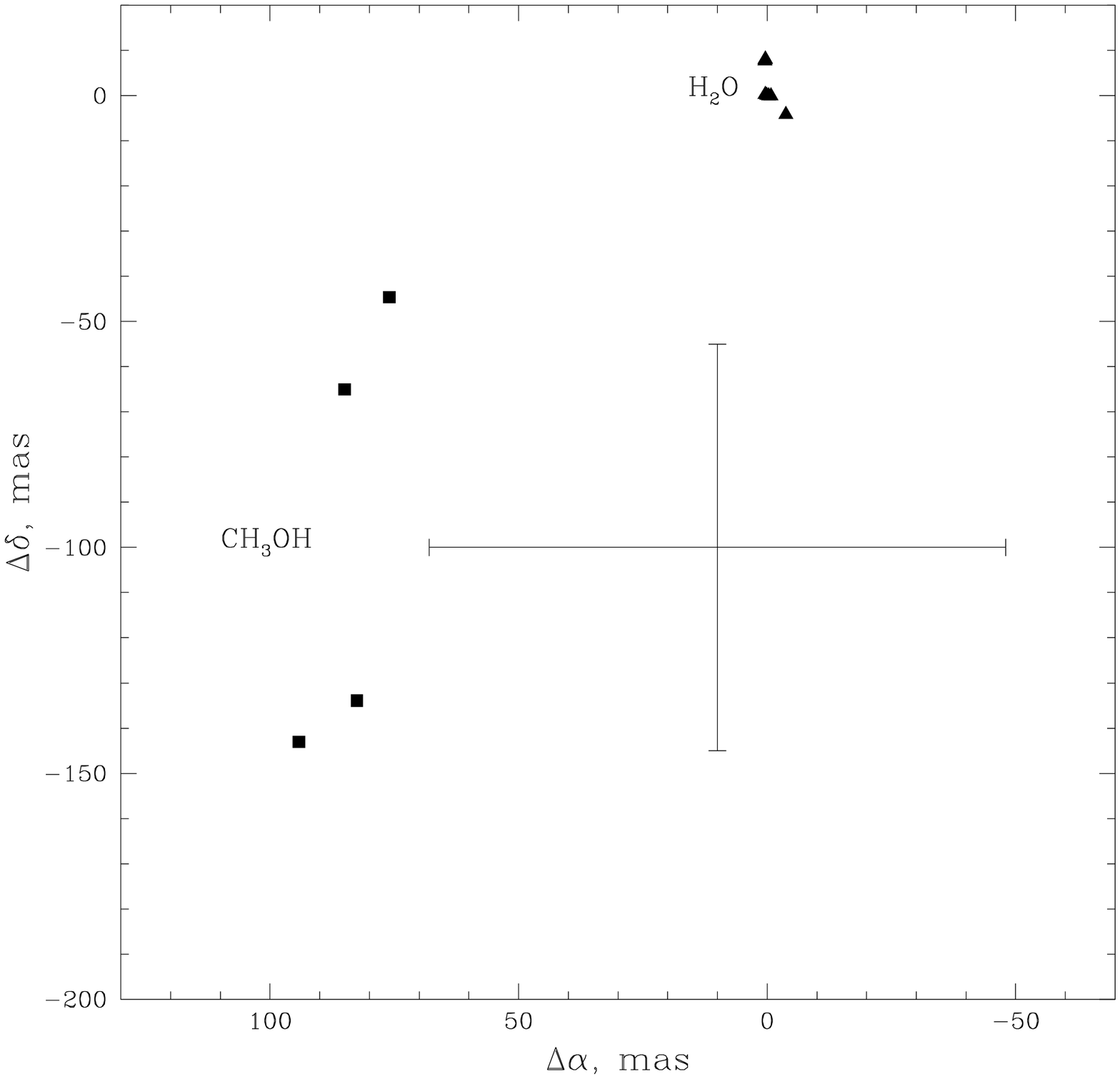}
\caption{The left figure is the map of the H$_2$O maser. The brightness of
every point is
a maximum from the brightnesses in the maps in every spectral channels
at every definite point. The right figure shows
relative position of methanol and H$_2$O maser spots. Positional uncertainty
between H$_2$O and methanol masers is indicated by cross bars.}
\end{figure}

Linear structure of   H$_2$O  and methanol masers can be 
modeled by the disk rotating around a central protostar N0. 
If the mass of central young star (or protostar) is 10 M$_\odot$  
and the distance is 6 kpc, H$_2$O maser components are 
located at the distance  40 AU from the star, and methanol 
maser components~-- at the distance (200 $-$ 800) AU. 
Physical objects responsible for the maser emission could 
be solid bodies with ice covered surfaces. The sublimation 
of ice composed of mixture of water and methanol, as in 
mantles of interstellar dust  particles and comets (Dartois 
at al. 1999, Shutte et al. 1996), could be a source of 
gaseous methanol and OH (from H$_2$O dissociation) in a 
shell around the solid bodies. The OH and methanol masers 
are generated in these shells. The proposed planets are 
rather similar to Edgeworth-Kuiper belt objects in the solar 
system (Slysh et al. 1999). 
So far as the thermal emission from the proposed 
planets would be strongly masked by the emission from 
hot dust in the compact HII region, it seems that, at 
present, there is no other means of detecting very distant 
planetary bodies except by their maser emission.

\begin{figure}
\plottwo{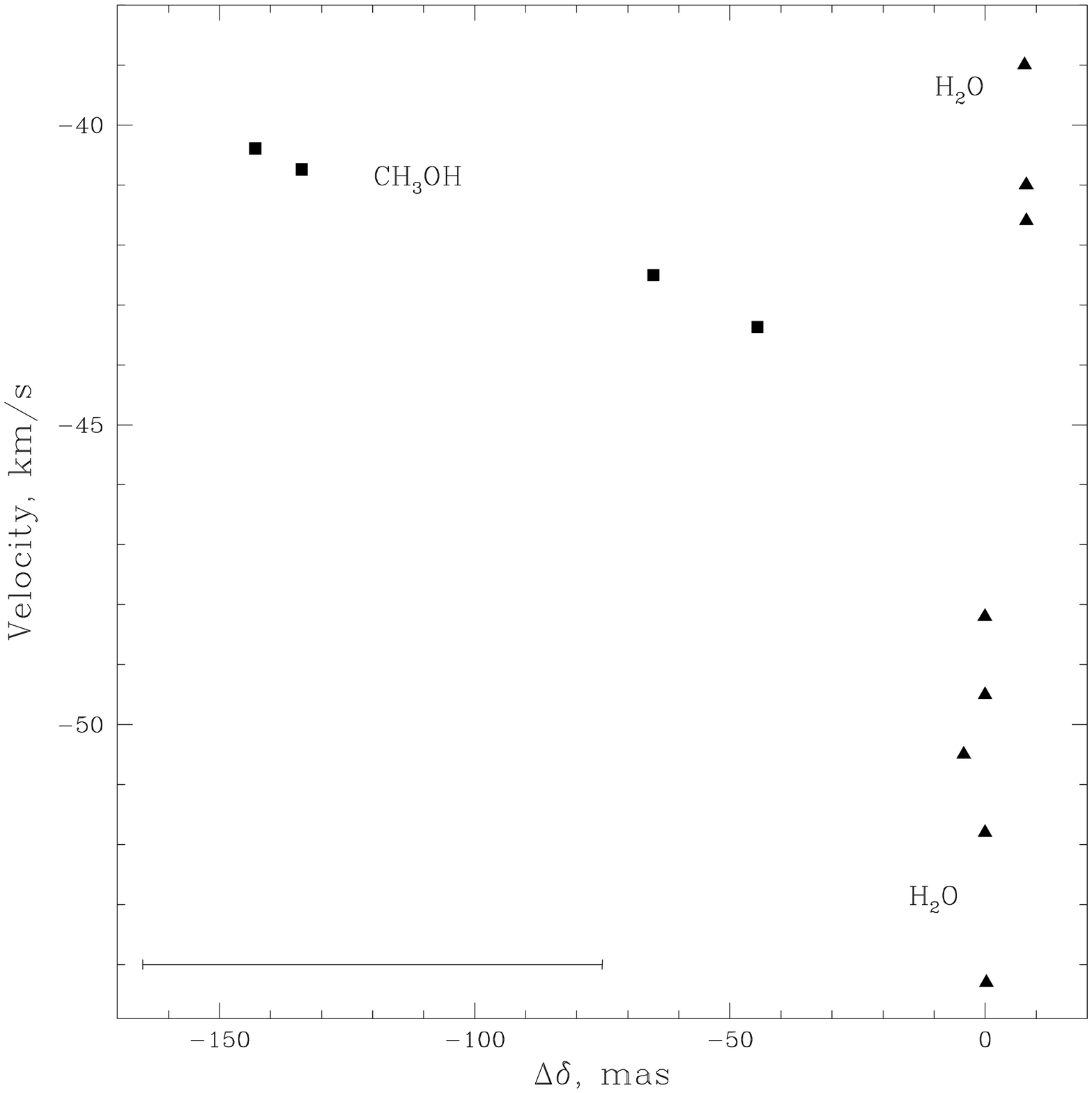}{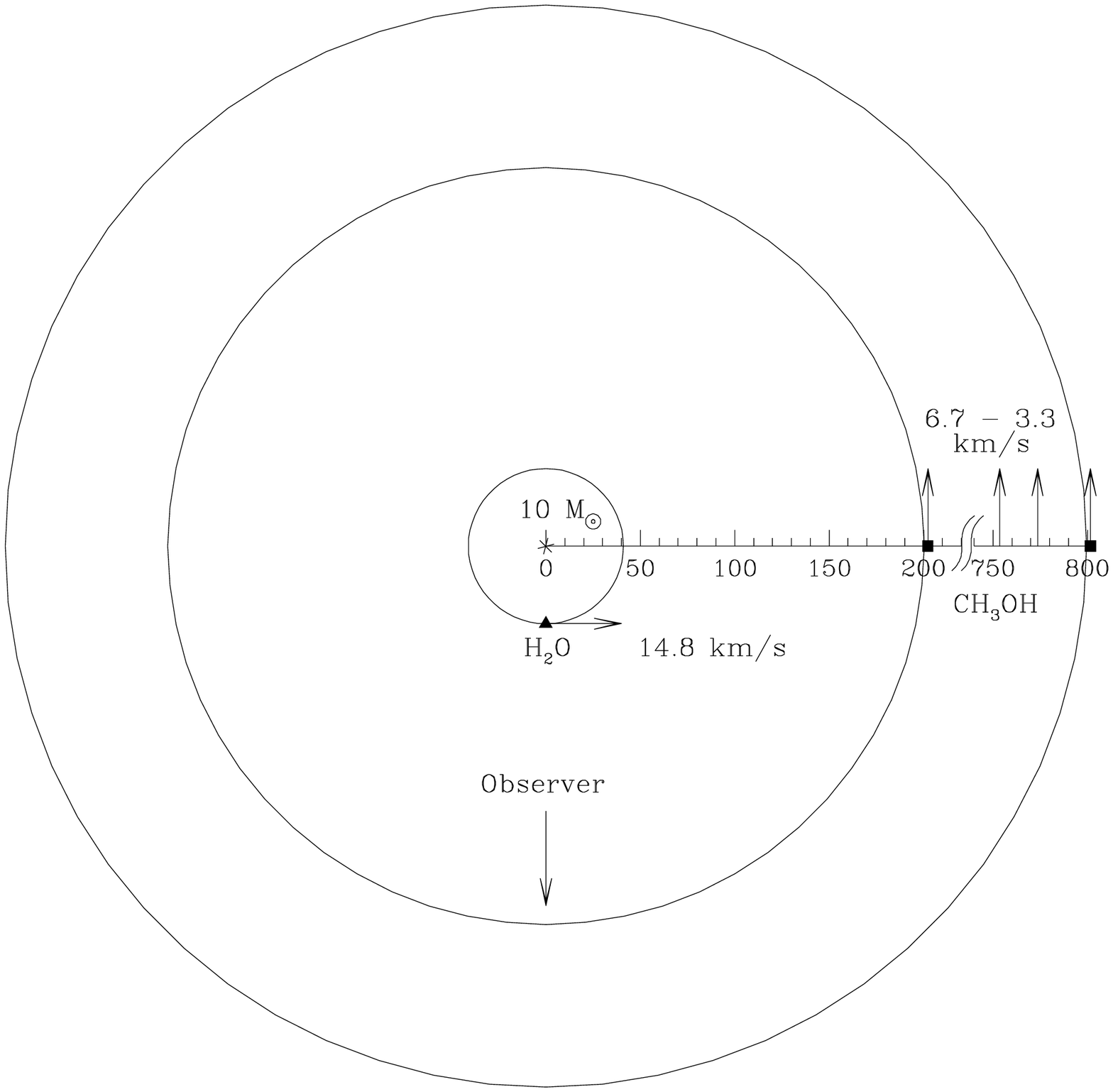}
\caption{The left figure is a declination~-- velocity diagram
for the maser spots in the direction North-South. In the left
lower corner the uncertainty of the relative position of H$_2$O and
methanol masers is shown. The right figure is a sketch of the disk
in GL2789. Numbers: distance from the star in a.u.;
arrows: direction and value of the orbital velocity of H$_2$O and
methanol masers.}
\end{figure}

\end{document}